\documentclass[twocolumn]{aastex62}

\usepackage{graphicx}	% Including figure files
\usepackage{amsmath}	% Advanced maths commands
\usepackage{amssymb}	% Extra maths symbols

\usepackage{bm}
\usepackage[shortlabels]{enumitem}

\received{January 1, 2018}
\revised{January 7, 2018}
\accepted{\today}
\submitjournal{ApJ}

\shorttitle{Coronal loop Kelvin-Helmholtz turbulence}
\shortauthors{A. Hillier et al}
\begin{document}

\title{Estimating the energy dissipation {from Kelvin-Helmholtz instability induced} turbulence in oscillating coronal loops}

\correspondingauthor{Andrew Hillier}
\email{a.s.hillier@exeter.ac.uk}

\author[0000-0002-0851-5362]{Andrew Hillier}
\affil{CEMPS, University of Exeter, Exeter EX4 4QF U.K.}

\author[0000-0000-0000-0000]{Tom Van Doorsselaere}
\affiliation{Centre for mathematical Plasma Astrophysics, Department of Mathematics, KU Leuven, Leuven, Belgium}

\author[0000-0001-5507-1891]{Konstantinos Karampelas}
\affiliation{Centre for mathematical Plasma Astrophysics, Department of Mathematics, KU Leuven, Leuven, Belgium}



\begin{abstract}
% Turbulent heating of coronal loops
% Prediction of the 
Kelvin-Helmholtz {instability induced} turbulence is one promising mechanism by which loops in the solar corona can be heated by MHD waves.
In this paper we present an analytical model of the dissipation rate of {Kelvin-Helmholtz instability induced} turbulence $\varepsilon_{\rm D}$, finding it scales as the wave amplitude ($d$) to the third power ($\varepsilon_{\rm D}\propto d^3$). Based on the concept of steady-state turbulence, we expect the turbulence heating throughout the volume of {the} loop to match the total energy injected through its footpoints. 
In situations where this holds, the wave amplitude has to vary as the cube-root of the injected energy. Comparing the analytic results with those of simulations shows that our analytic formulation captures the key aspects of the turbulent dissipation from the numerical work.
Applying this model to the observed characteristics of decayless kink waves we predict that the amplitudes of these observed waves is insufficient to turbulently heat the solar corona.
\end{abstract}



\section{Introduction}\label{INTRO}

Recent developments in the study of heating of coronal loops by MHD waves highlight the potential importance of turbulence forming at the boundary between the loop and the ambient corona \citep[e.g.][]{MAGYAR2016,KARA2017,KARAMPELAS2019a,KARAMPELAS2019b,HOW2017,TERR2018,ANT2017,ANT2018,HILL2019c}. This turbulence is created by the Kelvin-Helmholtz instability (KHi) \citep[e.g.][]{CHAN1961,HILL2019,BARB2019}, a shear-flow instability which can develop on the surface of oscillating flux tubes {\citep[see, for example,][]{Hayvaerts1983, Hollweg1987, Ofman1994GeoRL,SOLER2010, TERR2008, ANT2014,ANT2015, ANT2016}}, driven by the shear flows associated with the MHD kink wave. Once the turbulence has developed, it is the formation of small-scales in the velocity and magnetic fields which allows for fast dissipation of the wave energy making it a process that is possibly relevant to heat coronal loops.

In recent years, it has been observed that there are low-amplitude, transverse waves occurring in coronal loops. They were first detected in imaging data by \citet{wang2012}, and then later spectroscopically by \citet{tian2012}. They were shown to be different from the classic, impulsively excited transverse kink waves by \citet{nistico2013}, who named these waves {\em decayless waves}. \citet{ANFINOGENTOV2015} showed that these decayless waves are truly omnipresent, by selecting loops in subsequent active regions and showing that all these active regions show the decayless waves. The fact that these waves are omnipresent makes them an excellent candidate for heating the solar corona.

While the true mechanism for the existence of these decayless waves is still debated, they have been modelled numerically by \citep{KARA2017} as footpoint driven coronal loops, recovering many observational characteristics \citep{vd2018,guo2019a}. {Driven Alfv\'{e}n waves in inhomogeneous plasma have already been connected to coronal heating through the energy deposition at the resonant layers of {coronal loops}, by past numerical studies \citep{Poedts1989SoPh,Steinolfson1993ApJ}, while also establishing the development of KHi due to the strong shear velocities in those layers \citep{Ofman1994GeoRL,poedts1996}. Chromospheric coupling has been shown to lead to movement of those layers across the loop, resulting in heating in the entire loop volume \citep{Ofman1998ApJ}.} In the newer numerical studies, the driven oscillations are shown to develop long-lived turbulence throughout the loop providing continuous deposition of energy \citep{KARAMPELAS2019a,KARAMPELAS2019b} throughout its entire cross-section \citep{KARAMPELAS2018}. 
This can be seen as an example of quasi-steady-state turbulence, where the energy injected as large-scale oscillatory motions through the loop footpoints cascades to smaller scales through the KHi which is followed by energy dissipation once viscous or diffusive scales are reached.
A key question regarding this process is: how does the rate at which wave energy is dissipated depend on the amplitude of the oscillation of the loop structure? Indeed, this question is key in determining if the KHi is important in heating coronal loops, because it would allow for an observational estimate of the energy dissipation rate from the observed amplitudes of decayless waves \citep{ANFINOGENTOV2015}, which are accurately modelled with these driven loop simulations. 

The recent study of \citet{HILL2019c} highlighted how simple mean-field solutions could be developed for a Kelvin-Helmholtz mixing layer, and then applied to impulsively-excited oscillating prominence threads or coronal loops to constrain the energy available for heating.
In this paper, we adapt the solution of \citet{HILL2019c} to be applicable to driven oscillations, and use this to predict the wave amplitude dependence on injected energy flux for the case where the turbulent dissipation balances with the injected energy.
We then compare these predictions with the simulation results of \citet{KARAMPELAS2019a} and \citet{KARAMPELAS2019b} for validation of these results.

\section{Modelling}\label{model}

The philosophy we use to develop our model is very simple. We determine the energy contained in the turbulence of a mixing layer with fully developed KHi, and the timescales associated with this turbulence \citep[these are taken from][]{HILL2019c}. This is then used to estimate the rate at which energy is transferred between different spatial scales {perpendicular to the magnetic field} ($\varepsilon_{\rm T}$), a parameter that is often called ``energy cascade/dissipation rate'' in turbulence studies \citep[e.g.][]{vanderholst2014}. In steady state turbulence all the energy injected into the system cascades down through spatial scales (at the same rate it is injected), and then is dissipated {(again at the same rate it is injected)} \citep[e.g.][]{Yokoi2020}. Therefore, if one of the energy rates of the system can be determined, then they all can be determined.
Once we have used the results of \citet{HILL2019c} to develop the model for $\varepsilon_{\rm T}$ this will then be used to investigate the following:
\begin{enumerate}
\item For a given wave amplitude, what corresponding heating rate do we predict the loop produces?

\item For a given energy injection rate into a loop, what wave amplitude does the loop need to show in order to balance all the injected energy with turbulent dissipation?
\end{enumerate}

The first step is to estimate the energy dissipation rate $\varepsilon_{\rm D}$ (i.e. the expected heating rate). As explained above, in a turbulent layer which has reached a statistical steady-state, this will be the same as the energy transfer rate between scales ($\varepsilon_{\rm T}$). To approximate this from the results of \citet{HILL2019c} we use the mean turbulent kinetic energy of the KHi layer
\begin{equation}
\overline{\rm KE}_{\rm turb}\sim\frac{1}{4}\rho_{\rm mixed}\frac{\Delta V^2 (\alpha_1\alpha_2)^{1/2}}{(\sqrt{\alpha_1}+\sqrt{\alpha_2})^2},
\end{equation}
and the mixing timescale
\begin{equation}\label{mix_time}
\tau_{\rm mixing}\approx\frac{2l}{V_{\rm turb, RMS}}\ge\frac{2l}{\Delta V }\sqrt{2}\frac{\sqrt{\alpha_1}+\sqrt{\alpha_2}}{(\alpha_1\alpha_2)^{1/4}}.
\end{equation}
Here $1$ and $2$ are used to denote the values on either side of the mixing layer, for instance the density in the mixed layer $\rho_{\rm mixed}$ is written in terms of the density on either side of the mixing layer $\rho_{\rm mixed}=\sqrt{\rho_1\rho_2}$, the relative density is $\alpha_{1,2}=\rho_{1,2}/(\rho_{1}+\rho_2)$. We take $\Delta V$ as the velocity difference across the mixing layer, and $l$ is the layer half-width which is used to approximate the radius of the turbulent eddies. 
By dividing $\overline{\rm KE}_{\rm turb}$ by $\tau_{\rm mixing}$ we approximate $\varepsilon_{\rm T}$ to be:
\begin{equation}\label{epsilon}
\varepsilon_{\rm T} \approx \frac{\overline{\rm KE}_{\rm turb}}{\tau_{\rm mixing}}\lessapprox \frac{1}{4\sqrt{2}}\rho_{\rm mixed}\frac{\Delta V^3 (\alpha_1\alpha_2)^{3/4}}{2l(\sqrt{\alpha_1}+\sqrt{\alpha_2})^3}.
\end{equation}

Equation \ref{epsilon} shows how, for a given wave amplitude (which determines the velocity difference) and for given densities both inside and out of the tube, the dissipation in a mixing layer of width $2l$ can be estimated. Since the simulated flux tubes are shown to be fully mixed \citep{KARAMPELAS2018}, the thickness of the mixing layer, and with that the diameter of the turbulent eddies, can be estimated as the radius of the flux tube $R$, and the dissipation rate can be determined to be:
\begin{equation}\label{epsilon2}
\varepsilon_{\rm D}=\varepsilon_{\rm T} \lessapprox \frac{1}{4\sqrt{2}}\rho_{\rm mixed}\frac{\Delta V^3 (\alpha_1\alpha_2)^{3/4}}{R(\sqrt{\alpha_1}+\sqrt{\alpha_2})^3}.
\end{equation}

To use this to make any estimate of the heating rate in the solar corona, we have to understand how $\Delta V$ used in Equation \ref{epsilon2} relates to the wave velocity amplitude $V_{\rm AMP}$.
We determine that the magnitude of $\Delta V$ behaves as 
\begin{equation}
    \Delta V=\frac{2}{2^{3/2}}V_{\rm AMP},
\end{equation}
where the factor of $2$ in the nominator signifies the peak velocity shear is twice the velocity amplitude of the wave, and the factor of $2^{3/2}$ in the denominator relates to the fact we use the root-mean-squared velocity taking averages over one wave period, along the length of the tube and azimuthally around the tube.

Taking a coronal density of $0.601\times 10^{-15}$\,g\,cm$^{-3}$ and a loop density of $0.947\times 10^{-15}$\,g\,cm$^{-3}$, equal to the weighted mean densities of stratified loops \citep{ANDRIES2005} from the simulations of \citet{KARAMPELAS2019a}, a velocity amplitude of $V_{\rm AMP}=4\times 10^{6}$\,cm\,s$^{-1}$ (consistent with a wave amplitude of $\sim 10^8$\,cm and a period of $170$\,s), and a loop radius of $10^8$\,cm, this gives us a predicted $\varepsilon_{\rm D}$ of $\varepsilon_{\rm D}\approx 3.7 \times 10^{-6}$\,erg\,cm$^{-3}$\,s$^{-1}$. 
The rate at which the energy would be lost through radiative losses ($\varepsilon_{\rm RL}$) from the mixing layer is given by $\varepsilon_{\rm RL}=n_{\rm mixed}^2\Lambda(T)$ with $n$ the electron number density given by mixing ($n_{\rm mixed}=\sqrt{n_1n_2}$) and $\Lambda(T)$ the optical thin radiative loss function.
The numbers we use for our heating rate estimate are approximately equivalent to $n_{\rm mixed}=4.5\times 10^8$\,cm$^{-3}$ and $\Lambda(10^6K)\approx10^{-22}$\,erg\,cm$^3$\,s$^{-1}$ \citep{ANZER2008} which give $\varepsilon_{\rm RL}\approx 2 \times 10^{-5}$\,erg\,cm$^{-3}$\,s$^{-1}$.
This is a factor of $\sim5$ larger than our predicted heating rate.

Our next question is, if a system is undergoing statistically-steady turbulence, what is the wave amplitude that is required to give sufficiently strong turbulence so that all the injected wave energy is dissipated by the turbulence?
That is to say, for a given set of loop parameters and energy injection rate, there must be a $\Delta V$ which results in the wave-driven turbulence being sufficiently vigorous for this to happen.
Balancing the injected wave energy flux $E_{\rm FLUX}$ injected into the tube at both its footpoints with the dissipation throughout the loop volume leads to:
\begin{align}
2E_{\rm FLUX}\pi R^2 =& \pi R^2 L \varepsilon_{\rm D} \label{energyflux}\\ \approx &\frac{\pi R^2 L}{4\sqrt{2}}\rho_{\rm mixed}\left(\frac{V_{\rm AMP}}{\sqrt{2}}\right)^3\frac{(\alpha_1\alpha_2)^{3/4}}{R(\sqrt{\alpha_1}+\sqrt{\alpha_2})^3}\nonumber
\end{align}
where $L$ is the length of the loop and the factor of $2$ on the LHS is used to highlight that energy is injected via both footpoints.
This can be rearranged to solve for $V_{\rm AMP}$, {giving:
\begin{equation}\label{amp_vs_flux}
V_{\rm AMP}\approx 2^{5/3}\left(\frac{E_{\rm FLUX} R}{\rho_{\rm mixed}L} \right)^{1/3}\frac{(\sqrt{\alpha_1}+\sqrt{\alpha_2})}{(\alpha_1\alpha_2)^{1/4}}.
\end{equation}
Of} note here is that the wave amplitude is expected to scale as the cube-root of the injected energy flux.

{The connection between the velocity amplitude and the energy flux predicted here is fundamentally different from that coming from the WKB approximation where the energy flux would scale as
\begin{equation}
    E_{\rm FLUX}\propto V_{\rm AMP}^2V_{\rm KINK}
\end{equation}
where $V_{\rm KINK}$ is the kink speed \citep[e.g.][]{vd2014}. 
{However, as noted in \citet{KARAMPELAS2019b}, when a loop is driven at its resonant frequency, the amplitude of the wave can no longer be predicted by the WKB approximation. {Even though WKB theory gives the energy flux into the loop based on the velocity of the footpoint motions}, the resonance in the system results in energy getting trapped and the amplitude of the wave {increasing beyond the amplitude of the footpoint motions}.} This continues until nonlinearities or dissipative processes saturate its growth.
Even when a magnetic field is driven over a wide range of frequencies, for example as a result of being driven by convection, the tendency is for the wave energy to accumulate in the system at resonant frequencies \citep[e.g.][]{MATSUMOTO2010,afanasyev2020}.
Our model shows how you can connect the energy flux to the velocity amplitude of the system when a kink wave is driven at a resonant frequency and is nonlinearly saturated by the development of turbulence created by the Kelvin-Helmholtz instability.
}

\section{Comparison with simulation results}
To provide benchmarking of the estimates presented in Section \ref{model}, it is important to compare them with simulations. Here we use the results presented of simulations of driven oscillations in \citet{KARAMPELAS2019a} and \citet{KARAMPELAS2019b} as a way of benchmarking our model. 
We will focus on two of the results from these studies:
\begin{itemize}
\item From \citet{KARAMPELAS2019a} Figure 12, the slope of the internal energy increase as a result of the turbulent heating.

\item From \citet{KARAMPELAS2019b} Figure 6, the average amplitudes of the kink oscillations for different energy injection rates.
\end{itemize}

The main setup in both studies is a straight flux tube of radius $10^8$\,cm and length $2\times10^{10}$\,cm, consisting of gravitationally stratified plasma. The coronal background density at the footpoint is $0.836\times 10^{-15}$\,g\,cm$^{-3}$, three times lower than the loop density at the footpoint. A straight magnetic field of $B_z = 22.8$ G is considered, while the temperature varies across the tube axis (on the xy-plane), from $0.9$ MK inside the loop to $2.7$ MK outside. The models were allowed to reach a quasi-equilibrium state before introducing the driver. All calculations in the models considered here were performed in ideal MHD in the presence of numerical dissipation. {The excessive numerical dissipation compared to the physical dissipation does not greatly impact the results related to energy, because the energy cascade rate is determined at the larger scales by the turbulence properties. The power law dependence sets the flow of energy from large scale to small scale, where it is dissipated by dissipative processes. However, the specific dissipation process does not influence the energy flow down the scales in the part where ideal turbulence dominates.} 

The setups in \citet{KARAMPELAS2019a} have a resolution of $[15.63,15.63,1563]\times 10^5$\,cm  in the x, y and z direction, while a coarser grid of $[40,40,1563]\times 10^5$\,cm  was considered in \citet{KARAMPELAS2019b}. An important point about these simulations is that the wave driver period is set to that of the fundamental kink mode of the loop, so the energy of the driver can easily be trapped in the loop.

Looking at the slope of the internal energy increase in Figure 12 of \citet{KARAMPELAS2019a} and the rate of increase of internal energy (i.e. $\varepsilon_{\rm D}$)  can be estimated to be $\sim 2.3 \times 10^{-6}$\,erg\,cm$^{-3}$\,s$^{-1}$, when the whole box is taken into account, or $\sim 4.88 \times 10^{-6}$\,erg\,cm$^{-3}$\,s$^{-1}$, when only a region of radius $3 \times 10^8$\, cm containing the loop is considered. In conjunction with this, the estimate of the heating rate using Equation \ref{epsilon2} (made using appropriate parameters to match with the model of \citet{KARAMPELAS2019a}) we find a dissipation rate of $3.7\times 10^{-6}$\,erg\,cm$^{-3}$\,s$^{-1}$.
{Not only does this estimate give the same order of magnitude, but it is within a factor of less than 2 of the heating measured in the simulation}.
The heating in the simulations was found to be only 67\% efficient \citep{KARAMPELAS2019a}, implying that our model underestimates the 100\% efficiency heating rate by a factor of 2. 

\begin{figure}
    \centering
    \includegraphics[width=0.45\textwidth]{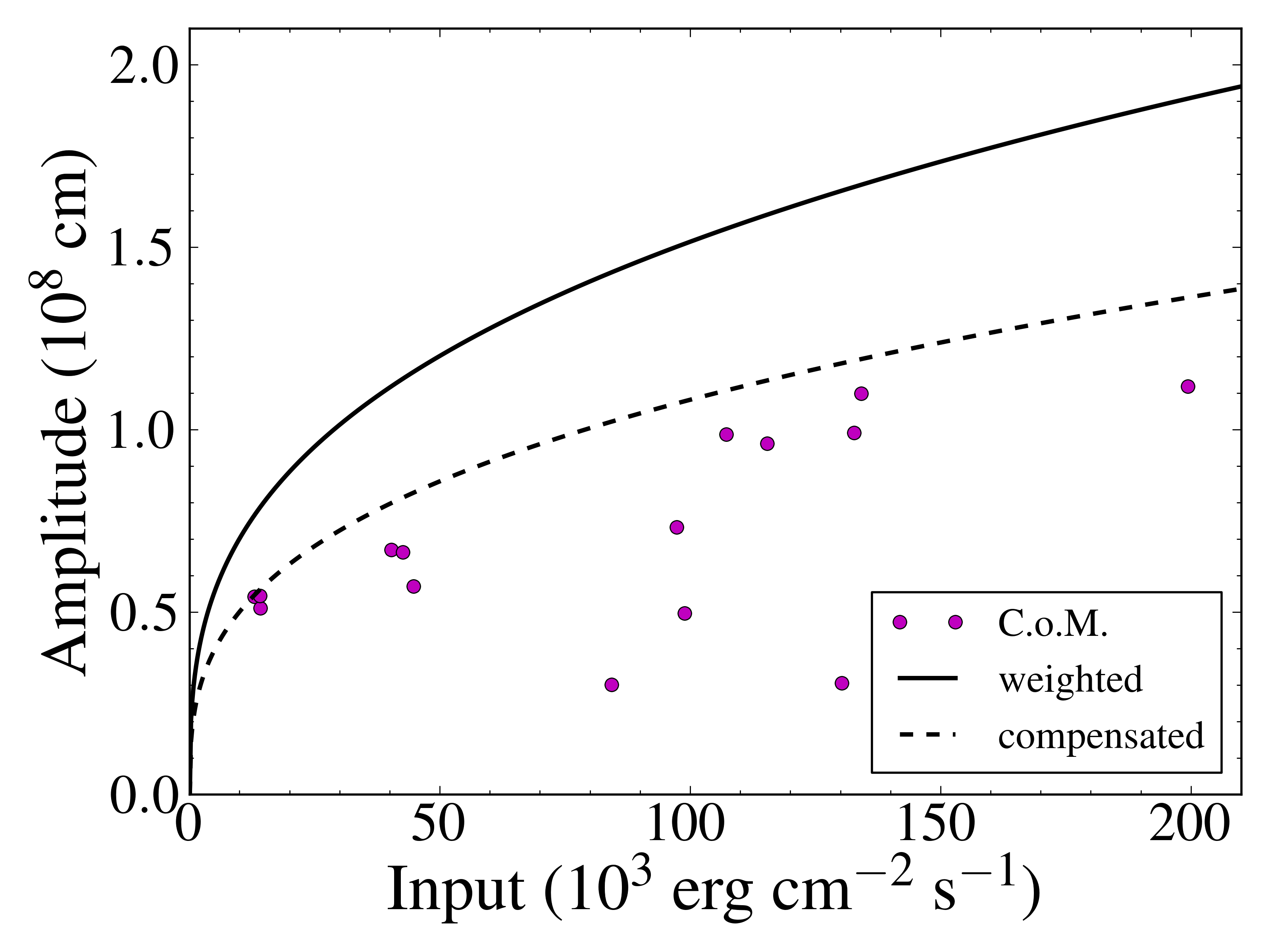}
    \caption{Plot of wave amplitude against input energy flux for the simulations presented in \citet{KARAMPELAS2019b} (purple dots) and the predicted amplitudes calculated using Equation \ref{amp_vs_flux} (solid line). The dashed line gives the prediction assuming the heating is 36\%  efficient compared to the model.}
    \label{wave_amp}
\end{figure}

Now that we have shown we can provide a good approximation of $\varepsilon_{\rm D}$, we turn to understanding how the wave amplitude could depend on the dissipation rate as a result of KHi turbulence. Figure \ref{wave_amp} shows the relation between the injected energy flux and the wave amplitude, where the wave amplitude $d$ is given by $d=V_{\rm AMP}/\omega$, with $\omega$ the wave frequency and $V_{\rm AMP}$ as given by Equation \ref{amp_vs_flux}. The parameters used were those applied in the simulations in \citet{KARAMPELAS2019b} including a weighted mean coronal density of $0.601\times 10^{-15}$\,g\,cm$^{-3}$ and a weighted mean loop density of $0.947\times 10^{-15}$\,g\,cm$^{-3}$ for our stratified loops \citep{ANDRIES2005}. Also plotted are the results from simulations of kink waves presented in \citet{KARAMPELAS2019b}. Here we show the average centre of mass (C.o.M.) displacement as a function of time, from the respective models.
The multiple points for a given driving energy correspond to different times, which implies a true steady state is never reached in the simulations presenting a deviation from the assumptions made.
We included a dashed line that compensates the efficiency of the model to match to the reduced efficiency of the heating from the simulations due to the smaller wave amplitudes compared to those found in \citet{KARAMPELAS2019a}.
It is clear that the curve determined by compensated Equation \ref{amp_vs_flux}, acts as an upper limit of the wave amplitudes measured from the simulations.

\section{Summary and Discussion}

There are two main conclusions from this work:
\begin{enumerate}
\item a good estimate for the energy dissipation as a result of turbulence driven by MHD kink waves in the solar corona can be calculated by extending the work of \citet{HILL2019c}.
\item This implies that the wave amplitude is proportional to the cube-root of the energy injection rate.
\end{enumerate}
{The corollary of all this is that the difference between dissipating enough energy via turbulence to balance radiative losses, and being an order of magnitude too small becomes only a difference in wave amplitude of $\sim 2.2$.}

Now we can also relate the results to the observational study of \citet{ANFINOGENTOV2015} {by working under the assumption that these are waves driven by noise with sufficient power at the resonant frequency of the system making the model presented here applicable}. There, they find that the average amplitude of the decayless oscillations is 0.17\,Mm. 
Taking also their average period of 258s, an assumed coronal density of $10^9$\,cm$^{-3}$, and a density contrast of $3$, we can apply our formula \ref{energyflux} to estimate the expected energy flux. 
{We find this to be $7.5\times 10^{-9}$\,erg\,cm$^{-3}$\,s$^{-1}$, much lower than the $3\times10^{-4}$\,erg\,cm$^{-3}$\,s$^{-1}$ which is expected for the optically thin radiative losses for $10^6$\,K plasma for the number densities stated above. 
Note that the big difference from the estimate in Section 2 comes from having a factor of $\sim 5$ decrease in the oscillation amplitude and a factor of $\sim1.5$ increase in the period, which corresponds to a total $\sim 7.5^3$ drop in heating rate.} 

In the above, we have always considered that the instability is driven by the large scale, observed velocity amplitude. However, it was shown recently by \citet{ANT2019} that resonant absorption plays a key role in the startup phase of the KHi. The Alfv\'en resonance will generate large, localised velocity gradients. So, our assumption that $\Delta V$ is of the same order of magnitude as $V_{\rm AMP}$ may be false if the resonance plays a big role. In that case, the $\Delta V$ could be increased by a factor that is much larger than 1. 
Along with this the radius of the eddies would shrink due to the localisation, which will further enhance the dissipation rate. One caveat to this is that even though the dissipation is faster, it happens in a {significantly smaller volume at the Reynolds numbers of the solar corona}, meaning that it is still necessary to transport this thermal energy throughout the rest of the loop. 
Along the magnetic field, field aligned thermal conduction can perform this role effectively, but across the field turbulent transport is likely to play the dominant role, the timescales for which are likely to be {longer} due to the large scales over which heat has to be transported.

{One point that is important to understand to give context to the discussion of the previous paragraph is that the KHi solution used here is one that is based on a discontinuous velocity field but is applicable once a broad turbulent layer has formed and is dynamically determined by the total energy originally contained within the layer. Even with resonant absorption occurring, the total energy contained over the broad mixing layer is unlikely to be significantly larger than the value we use in our model based on the wave amplitude \citep[this argument is borne out by the accurate prediction by our model of the heating rate found in][]{KARAMPELAS2019a}. Looking at the simulations of \citet{MAGYAR2016} for an initially discontinuous tube boundary (close to the basis of our model), and a smooth boundary where resonant absorption can play a key role, larger amplitude perturbations only show small qualitative and quantitative differences, meaning our model will be applicable even when resonant absorption is playing an important role in the initialisation of the Kelvin-Helmholtz dynamics. 
However, the larger differences found in that study for much smaller amplitude waves mean that the role of resonant absorption in the development of turbulence merits further study in terms of understanding the turbulent heating from decayless oscillations, though this is beyond the scope of this study.}

{For the wave amplitude to increase until the dissipation rate matches the injection rate, energy must be driven into the loop at periods resonant with the fundamental or higher order harmonics of the kink mode of the loop.} Otherwise energy driven into the loop will be able to leak out instead of increasing the oscillation amplitude, therefore making it impossible for steady turbulence to develop.

\acknowledgments

AH was supported by his STFC Ernest Rutherford Fellowship grant number ST/L00397X/2 and STFC research grant ST/R000891/1. AH thanks the Europe Network Fund of the University of Exeter for providing support to visit KU Leuven. TVD was supported by the European Research Council (ERC) under the European Union's Horizon 2020 research and innovation programme (grant agreement No 724326) and the C1 grant TRACESpace of Internal Funds KU Leuven. KK is supported by a postdoctoral mandate from KU Leuven Internal Funds (PDM/2019).




\begin{thebibliography}{}
\expandafter\ifx\csname natexlab\endcsname\relax\def\natexlab#1{#1}\fi
\providecommand{\url}[1]{\href{#1}{#1}}
\providecommand{\dodoi}[1]{doi:~\href{http://doi.org/#1}{\nolinkurl{#1}}}
\providecommand{\doeprint}[1]{\href{http://ascl.net/#1}{\nolinkurl{http://ascl.net/#1}}}
\providecommand{\doarXiv}[1]{\href{https://arxiv.org/abs/#1}{\nolinkurl{https://arxiv.org/abs/#1}}}

\bibitem[{{Afanasyev} {et~al.}(2020){Afanasyev}, {Van Doorsselaere}, \&
  {Nakariakov}}]{afanasyev2020}
{Afanasyev}, A.~N., {Van Doorsselaere}, T., \& {Nakariakov}, V.~M. 2020, \aap,
  633, L8, \dodoi{10.1051/0004-6361/201937187}

\bibitem[{{Andries} {et~al.}(2005){Andries}, {Goossens}, {Hollweg}, {Arregui},
  \& {Van Doorsselaere}}]{ANDRIES2005}
{Andries}, J., {Goossens}, M., {Hollweg}, J.~V., {Arregui}, I., \& {Van
  Doorsselaere}, T. 2005, \aap, 430, 1109, \dodoi{10.1051/0004-6361:20041832}

\bibitem[{{Anfinogentov} {et~al.}(2015){Anfinogentov}, {Nakariakov}, \&
  {Nistic{\`o}}}]{ANFINOGENTOV2015}
{Anfinogentov}, S.~A., {Nakariakov}, V.~M., \& {Nistic{\`o}}, G. 2015, \aap,
  583, A136, \dodoi{10.1051/0004-6361/201526195}

\bibitem[{{Antolin} {et~al.}(2016){Antolin}, {De Moortel}, {Van Doorsselaere},
  \& {Yokoyama}}]{ANT2016}
{Antolin}, P., {De Moortel}, I., {Van Doorsselaere}, T., \& {Yokoyama}, T.
  2016, \apj, 830, L22, \dodoi{10.3847/2041-8205/830/2/L22}

\bibitem[{{Antolin} {et~al.}(2017){Antolin}, {De Moortel}, {Van Doorsselaere},
  \& {Yokoyama}}]{ANT2017}
---. 2017, \apj, 836, 219, \dodoi{10.3847/1538-4357/aa5eb2}

\bibitem[{{Antolin} {et~al.}(2015){Antolin}, {Okamoto}, {De Pontieu},
  {Uitenbroek}, {Van Doorsselaere}, \& {Yokoyama}}]{ANT2015}
{Antolin}, P., {Okamoto}, T.~J., {De Pontieu}, B., {et~al.} 2015, \apj, 809,
  72, \dodoi{10.1088/0004-637X/809/1/72}

\bibitem[{{Antolin} {et~al.}(2018){Antolin}, {Schmit}, {Pereira}, {De Pontieu},
  \& {De Moortel}}]{ANT2018}
{Antolin}, P., {Schmit}, D., {Pereira}, T.~M.~D., {De Pontieu}, B., \& {De
  Moortel}, I. 2018, \apj, 856, 44, \dodoi{10.3847/1538-4357/aab34f}

\bibitem[{Antolin \& Van~Doorsselaere(2019)}]{ANT2019}
Antolin, P., \& Van~Doorsselaere, T. 2019, Frontiers in Physics, 7, 85,
  \dodoi{10.3389/fphy.2019.00085}

\bibitem[{{Antolin} {et~al.}(2014){Antolin}, {Yokoyama}, \& {Van
  Doorsselaere}}]{ANT2014}
{Antolin}, P., {Yokoyama}, T., \& {Van Doorsselaere}, T. 2014, \apj, 787, L22,
  \dodoi{10.1088/2041-8205/787/2/L22}

\bibitem[{{Anzer} \& {Heinzel}(2008)}]{ANZER2008}
{Anzer}, U., \& {Heinzel}, P. 2008, \aap, 480, 537,
  \dodoi{10.1051/0004-6361:20078832}

\bibitem[{{Barbulescu} {et~al.}(2019){Barbulescu}, {Ruderman}, {Van
  Doorsselaere}, \& {Erd{\'e}lyi}}]{BARB2019}
{Barbulescu}, M., {Ruderman}, M.~S., {Van Doorsselaere}, T., \& {Erd{\'e}lyi},
  R. 2019, \apj, 870, 108, \dodoi{10.3847/1538-4357/aaf506}

\bibitem[{{Chandrasekhar}(1961)}]{CHAN1961}
{Chandrasekhar}, S. 1961, {Hydrodynamic and hydromagnetic stability}
  ({Clarendon Press})

\bibitem[{{Guo} {et~al.}(2019){Guo}, {Van Doorsselaere}, {Karampelas}, {Li},
  {Antolin}, \& {De Moortel}}]{guo2019a}
{Guo}, M., {Van Doorsselaere}, T., {Karampelas}, K., {et~al.} 2019, \apj, 870,
  55, \dodoi{10.3847/1538-4357/aaf1d0}

\bibitem[{{Heyvaerts} \& {Priest}(1983)}]{Hayvaerts1983}
{Heyvaerts}, J., \& {Priest}, E.~R. 1983, \aap, 117, 220

\bibitem[{{Hillier} \& {Arregui}(2019)}]{HILL2019c}
{Hillier}, A., \& {Arregui}, I. 2019, \apj, 885, 101,
  \dodoi{10.3847/1538-4357/ab4795}

\bibitem[{{Hillier} {et~al.}(2019){Hillier}, {Barker}, {Arregui}, \&
  {Latter}}]{HILL2019}
{Hillier}, A., {Barker}, A., {Arregui}, I., \& {Latter}, H. 2019, \mnras, 482,
  1143, \dodoi{10.1093/mnras/sty2742}

\bibitem[{{Hollweg}(1987)}]{Hollweg1987}
{Hollweg}, J.~V. 1987, \apj, 317, 918, \dodoi{10.1086/165341}

\bibitem[{{Howson} {et~al.}(2017){Howson}, {De Moortel}, \&
  {Antolin}}]{HOW2017}
{Howson}, T.~A., {De Moortel}, I., \& {Antolin}, P. 2017, \aap, 607, A77,
  \dodoi{10.1051/0004-6361/201731178}

\bibitem[{{Karampelas} \& {Van Doorsselaere}(2018)}]{KARAMPELAS2018}
{Karampelas}, K., \& {Van Doorsselaere}, T. 2018, \aap, 610, L9,
  \dodoi{10.1051/0004-6361/201731646}

\bibitem[{{Karampelas} {et~al.}(2017){Karampelas}, {Van Doorsselaere}, \&
  {Antolin}}]{KARA2017}
{Karampelas}, K., {Van Doorsselaere}, T., \& {Antolin}, P. 2017, \aap, 604,
  A130, \dodoi{10.1051/0004-6361/201730598}

\bibitem[{{Karampelas} {et~al.}(2019{\natexlab{a}}){Karampelas}, {Van
  Doorsselaere}, \& {Guo}}]{KARAMPELAS2019a}
{Karampelas}, K., {Van Doorsselaere}, T., \& {Guo}, M. 2019{\natexlab{a}},
  \aap, 623, A53, \dodoi{10.1051/0004-6361/201834309}

\bibitem[{{Karampelas} {et~al.}(2019{\natexlab{b}}){Karampelas}, {Van
  Doorsselaere}, {Pascoe}, {Guo}, \& {Antolin}}]{KARAMPELAS2019b}
{Karampelas}, K., {Van Doorsselaere}, T., {Pascoe}, D.~J., {Guo}, M., \&
  {Antolin}, P. 2019{\natexlab{b}}, Frontiers in Astronomy and Space Sciences,
  6, 38, \dodoi{10.3389/fspas.2019.00038}

\bibitem[{{Magyar} \& {Van Doorsselaere}(2016)}]{MAGYAR2016}
{Magyar}, N., \& {Van Doorsselaere}, T. 2016, \aap, 595, A81,
  \dodoi{10.1051/0004-6361/201629010}

\bibitem[{{Matsumoto} \& {Shibata}(2010)}]{MATSUMOTO2010}
{Matsumoto}, T., \& {Shibata}, K. 2010, \apj, 710, 1857,
  \dodoi{10.1088/0004-637X/710/2/1857}

\bibitem[{{Nistic{\`o}} {et~al.}(2013){Nistic{\`o}}, {Nakariakov}, \&
  {Verwichte}}]{nistico2013}
{Nistic{\`o}}, G., {Nakariakov}, V.~M., \& {Verwichte}, E. 2013, \aap, 552,
  A57, \dodoi{10.1051/0004-6361/201220676}

\bibitem[{{Ofman} {et~al.}(1994){Ofman}, {Davila}, \&
  {Steinolfson}}]{Ofman1994GeoRL}
{Ofman}, L., {Davila}, J.~M., \& {Steinolfson}, R.~S. 1994, \grl, 21, 2259,
  \dodoi{10.1029/94GL01416}

\bibitem[{{Ofman} {et~al.}(1998){Ofman}, {Klimchuk}, \&
  {Davila}}]{Ofman1998ApJ}
{Ofman}, L., {Klimchuk}, J.~A., \& {Davila}, J.~M. 1998, \apj, 493, 474,
  \dodoi{10.1086/305109}

\bibitem[{{Poedts} \& {Boynton}(1996)}]{poedts1996}
{Poedts}, S., \& {Boynton}, G.~C. 1996, \aap, 306, 610

\bibitem[{{Poedts} {et~al.}(1989){Poedts}, {Goossens}, \&
  {Kerner}}]{Poedts1989SoPh}
{Poedts}, S., {Goossens}, M., \& {Kerner}, W. 1989, \solphys, 123, 83,
  \dodoi{10.1007/BF00150014}

\bibitem[{{Soler} {et~al.}(2010){Soler}, {Terradas}, {Oliver}, {Ballester}, \&
  {Goossens}}]{SOLER2010}
{Soler}, R., {Terradas}, J., {Oliver}, R., {Ballester}, J.~L., \& {Goossens},
  M. 2010, \apj, 712, 875, \dodoi{10.1088/0004-637X/712/2/875}

\bibitem[{{Steinolfson} \& {Davila}(1993)}]{Steinolfson1993ApJ}
{Steinolfson}, R.~S., \& {Davila}, J.~M. 1993, \apj, 415, 354,
  \dodoi{10.1086/173169}

\bibitem[{{Terradas} {et~al.}(2008){Terradas}, {Andries}, {Goossens},
  {Arregui}, {Oliver}, \& {Ballester}}]{TERR2008}
{Terradas}, J., {Andries}, J., {Goossens}, M., {et~al.} 2008, \apj, 687, L115,
  \dodoi{10.1086/593203}

\bibitem[{{Terradas} {et~al.}(2018){Terradas}, {Magyar}, \& {Van
  Doorsselaere}}]{TERR2018}
{Terradas}, J., {Magyar}, N., \& {Van Doorsselaere}, T. 2018, \apj, 853, 35,
  \dodoi{10.3847/1538-4357/aa9d0f}

\bibitem[{{Tian} {et~al.}(2012){Tian}, {McIntosh}, {Wang}, {Ofman}, {De
  Pontieu}, {Innes}, \& {Peter}}]{tian2012}
{Tian}, H., {McIntosh}, S.~W., {Wang}, T., {et~al.} 2012, \apj, 759, 144,
  \dodoi{10.1088/0004-637X/759/2/144}

\bibitem[{{van der Holst} {et~al.}(2014){van der Holst}, {Sokolov}, {Meng},
  {Jin}, {Manchester}, {T{\'o}th}, \& {Gombosi}}]{vanderholst2014}
{van der Holst}, B., {Sokolov}, I.~V., {Meng}, X., {et~al.} 2014, \apj, 782,
  81, \dodoi{10.1088/0004-637X/782/2/81}

\bibitem[{{Van Doorsselaere} {et~al.}(2018){Van Doorsselaere}, {Antolin}, \&
  {Karampelas}}]{vd2018}
{Van Doorsselaere}, T., {Antolin}, P., \& {Karampelas}, K. 2018, \aap, 620,
  A65, \dodoi{10.1051/0004-6361/201834086}

\bibitem[{{Van Doorsselaere} {et~al.}(2014){Van Doorsselaere}, {Gijsen},
  {Andries}, \& {Verth}}]{vd2014}
{Van Doorsselaere}, T., {Gijsen}, S.~E., {Andries}, J., \& {Verth}, G. 2014,
  \apj, 795, 18, \dodoi{10.1088/0004-637X/795/1/18}

\bibitem[{{Wang} {et~al.}(2012){Wang}, {Ofman}, {Davila}, \& {Su}}]{wang2012}
{Wang}, T., {Ofman}, L., {Davila}, J.~M., \& {Su}, Y. 2012, \apjl, 751, L27,
  \dodoi{10.1088/2041-8205/751/2/L27}

\bibitem[{Yokoi(2020)}]{Yokoi2020}
Yokoi, N. 2020, Turbulence, Transport and Reconnection, ed. D.~MacTaggart \&
  A.~Hillier (Cham: Springer International Publishing), 177--265.
\newblock \url{https://doi.org/10.1007/978-3-030-16343-3_6}

\end{thebibliography}
\end{document}